\documentstyle[epsfig]{aipproc}

\begin{document}

\title{The Angular Momentum of Accreting Neutron Stars}
 
\author{Lars Bildsten} 
\address{Department of Physics and Department of Astronomy\\  University of
California, Berkeley, CA 94720}

\maketitle

\begin{abstract}
I review the rotation measurements of accreting neutron stars. Many of
the highly magnetic ($B>10^{11} \ {\rm G}$) accreting X-ray pulsars
have been continuously observed with the Burst and Transient
Source Experiment (BATSE) aboard the {\it Compton
Gamma-Ray Observatory} (CGRO) since April 1991. These observations
show that the accretion torque exerted on many disk-fed accreting
X-ray pulsars  changes sign on a monthly to yearly
timescale. This results in alternating periods of spin-up and
spin-down with nearly the same torques, leading to little net angular
momentum gained by accretion. I also summarize recent discoveries with
the {\it Rossi X-Ray Timing Explorer} (RXTE) of periodicities during
Type I X-ray bursts. These seem to indicate that many of the rapidly
accreting ($\dot M > 10^{-10} M_\odot \ {\rm yr}^{-1}$) and weakly
magnetic ($B\ll 10^{11} \ {\rm G}$) neutron stars in our galaxy are
rotating at frequencies $\gtrsim 250 \ {\rm Hz}$. Most remarkable is
that they all rotate within a rather narrow range of frequencies.
\end{abstract}

\section{Introduction}

Since the discovery of neutron stars (NSs) over 30 years ago,
there have been important open questions about the origin and
evolution of their rotation.  Measuring the NS rotation rates in the
low-mass X-ray binaries has been a long-standing goal of X-ray
astronomy and is motivated by the desire to find the
progenitors of the millisecond radio pulsars. 

There are now two methods of observing the spin in accreting NS. If
the NS has a misaligned magnetic dipole, then the brightness asymmetry
will be modulated by rotation. I will first review such observations
for the accreting X-ray pulsars. Even if the star is not magnetic,
modulations are possible during the rise (about one second) of a Type
I X-ray burst, as it is then that the burning front is traveling
around the star \cite{fry82,noz84,bil93,bil95}. This has provided a
new window on rotation for the weakly magnetic ($B \ll 10^{11} \ {\rm
G}$) NS.

 Before launching into a summary of the accreting NSs, let me briefly
summarize their less degenerate counterparts, the accreting white
dwarfs (WDs). Spin periods are directly measured for those WDs
sufficiently magnetized to channel the accretion flow onto their polar
caps. The most rapid WD rotator known to be magnetic is AE Aqr
($P_s\approx 33 $ s, see \cite{patt} \ for an overview), one of the
large class of DQ Her-type magnetic WD's that have accretion
disks. The magnetospheres for these objects are not much larger than
their radius and so their final spin periods are not all that
different than the dwarf novae (see below). The WD moment of inertia
is so large that measuring the accretion torque takes years, so that
there is typically just one torque measurement for each object. At
present, five are spinning-up and two are spinning down
\cite{patt}. Attempts to measure rotational broadening of photospheric
absorption lines from isolated DA WDs have been unsuccessful for
years, implying very slow rotation \cite{pil87,koes88,heb976}. This
agrees with the extremely slow rotation (5 hours to 2.5 days,
\cite{kep95}) measured from the frequency splitting of non-radial
oscillations in the ZZ Cetis.\footnote{ Unfortunately it is still the
case that no positive identification of a non-radial pulsation on an
accreting NS has been made (see \cite{bil98} for a brief summary).}
Rapid rotation has recently been inferred for accreting WDs in dwarf
novae systems, where presumably the magnetic field is low.
Observations of dwarf novae after outburst have found that the WD is
still hot for a time after the rapid accretion from the thermal disk
instability has halted (see \cite{sio95} for an overview).  This has
allowed the observers to see the WD photosphere in some detail and
yielded direct measurements of rotational line broadening, where the
largest value is for WZ Sagittae \cite{cheng97}, where $v_{rot} \sin
i=1200_{-400}^{+300} \ {\rm km/s}$. The others are VW Hyi, with
$v_{rot}\sin i\approx 600 \ {\rm km/s}$ \cite{sionetal95} and U Gem
with $v_{rot}\sin i\approx 100 \ {\rm km/s}$ \cite{sionetal94}. These
measurements imply WD rotation periods on the order of 30-300 seconds,
much faster than the non-accreting WD's.

The prolonged accretion of material is expected to drastically alter
the angular momentum of the NS. The accretion torque depends on the
amount of angular momentum brought in by the accreting matter. For
example, if it always arrives with the specific angular momentum of a
particle orbiting at the stellar radius ($R=10 \ {\rm km}$) it takes
only $\approx 10^7 \ {\rm yr}$ of accretion at $\dot M\approx 10^{-9}
M_\odot \ {\rm yr^{-1}}$ for a $M=1.4 M_\odot$ NS to reach
$\nu_s\approx 50$
Hz from an initially low frequency. As we now discuss, other physics
intervenes for the highly magnetic ($B >  10^{11} \ {\rm G}$) pulsars
so spin frequencies this high are only seen in the weakly magnetic NSs.

\section{Torques on the X-Ray Pulsars } 

The small moment of inertia and large torques allow for repeated
measurements of the accretion torques on X-ray pulsars.  Indeed,
BATSE's continuous observations have provided a monitor of the torque
exerted during magnetic accretion \cite{bil97}. These observations
have found that spin-up and spin-down are nearly equally prevalent in
these systems \cite{bil97,nel97}, contrary to the picture in the
1970's, when most accreting pulsars were thought to be spinning up
steadily (see Figure 5 in \cite{jossrap84}). BATSE has also tested
theories of accretion torque and magnetospheric QPO generation on
timescales of days \cite{fing96}, which we will not review here.

Let's start with some simple estimates that set the stage for the
discussion of the observations for the magnetic X-ray pulsars. In
these objects, the magnetic field cannot be ignored. The accreting
pulsar will experience a spin-up torque
\begin{equation}
N\approx \dot M(GMr_m)^{1/2},
\label{eq:torque}
\end{equation}
assuming the gas deposits its angular
momentum at the magnetospheric boundary, and that field lines
transport all of this angular momentum to the star
\cite{pring72,rapjos77}. The variable $r_m=\xi r_A$ is the
magnetospheric radius with
\begin{equation}
r_A\equiv\left(\mu^4\over {2GM\dot M^2}\right)^{1/7}=
6.8\times 10^8 \ {\rm cm}
\left({\mu\over 10^{30} \ {\rm G\ cm^3}}\right)^{4/7}
\left({10^{-10} \ M_\odot \ {\rm yr^{-1}}}
\over {\dot M}\right)^{2/7},
\end{equation}
being a characteristic length found by equating magnetic and fluid
stresses for a NS with magnetic dipole moment $\mu$. Estimates for the
dimensionless number $\xi$ range from $\approx 0.52$ \cite{gl79} to
$\approx 1$ \cite{aro93,ostr95,wang96}.  The detailed physics by which
material at this magnetospheric boundary loses its orbital angular
momentum, becomes entrained on the magnetic field lines, and makes its
way to the magnetic polar caps on the star is beyond this review.

Accretion will be inhibited by a centrifugal barrier if the pulsar
magnetosphere rotates faster than the Kepler frequency at the
magnetosphere, $r_m$. 
For accretion to continue, the magnetospheric radius
must lie inside the corotation radius defined by the NS spin period,
$P_s$, 
\begin{equation}
r_{co} = \left ({{GM} P_s^2 \over 4\pi^2} \right )^{1/3}
= 1.7 \times 10^8 \left(P_s\over {\rm s}\right)^{2/3} \rm{\; cm}, 
\label{r_co}
\end{equation}
so that if we demand that $r_m< r_{\rm co}$ there is 
a characteristic torque,
\begin{equation}
N_o\equiv \dot M(GM r_{\rm co})^{1/2},
\label{eq:maxtorque}
\end{equation}
which is convenient to use as it only depends on the NS spin period,
$P_s$, and $\dot M$. A pulsar subject to the torque in equation
(\ref{eq:torque}) will spin-up at a rate
\begin{equation}
\dot \nu = {N\over 2\pi I}= 1.6\times 10^{-13} {\rm s^{-2}}
\left({\dot M}\over {10^{-10} M_\odot \ {\rm yr^{-1}}}\right)
\left(P_s\over {\rm s}\right)^{1/3}
\left({r_m\over r_{co}}\right)^{1/2},
\label{eq:nudot}
\end{equation}
where $I\approx 0.4 M R^2$ is the NS's moment of inertia. 
The timescale for spinning up the NS is then
\begin{equation}
t_{\rm spin-up}\equiv -{\nu\over \dot \nu} \simeq 2\times 10^5 \ {\rm yr}
\left({10^{-10} M_\odot \  {\rm yr^{-1}}}\over \dot M\right)
\left({\rm s} \over P_s\right)^{4/3}
\left({r_{co}\over r_{m}}\right)^{1/2},
\label{eq:tspin}
\end{equation}
much shorter than the ages of most X-ray binaries \cite{els80}.
 Hence in this simple picture the NS gets spun up
until the spin frequency matches the Kepler frequency at the
magnetosphere (or where $r_m\approx r_{co}$)
\begin{equation}
P_{s,eq} \approx 8 \ {\rm s} \ \xi^{3/2} 
\left({10^{-10} M_\odot \ {\rm yr^{-1}}}\over 
{\dot M} \right)^{3/7}
\left({\mu\over 10^{30} \ {\rm G\ cm^{3}}}\right)^{6/7}.
\label{eq:peq}
\end{equation}
Presumably, NSs with $P_s< P_{s,eq}$ cannot easily accrete and may
experience a strong spin-down torque via the propeller effect
\cite{ill75}. Hence, none of the high field objects are rapidly
rotating. If it is true that the X-ray pulsars are at $P_s\sim
P_{s,eq}$, then one infers magnetic field strengths in the range
$10^{11} -10^{14} \ {\rm G}$.  The instantaneous $\dot M$ and torques
can be much different than their long-term averages, however, so one
only obtains a rough indication of $B$ this way.

\begin{figure}
\centerline{\epsfig{file=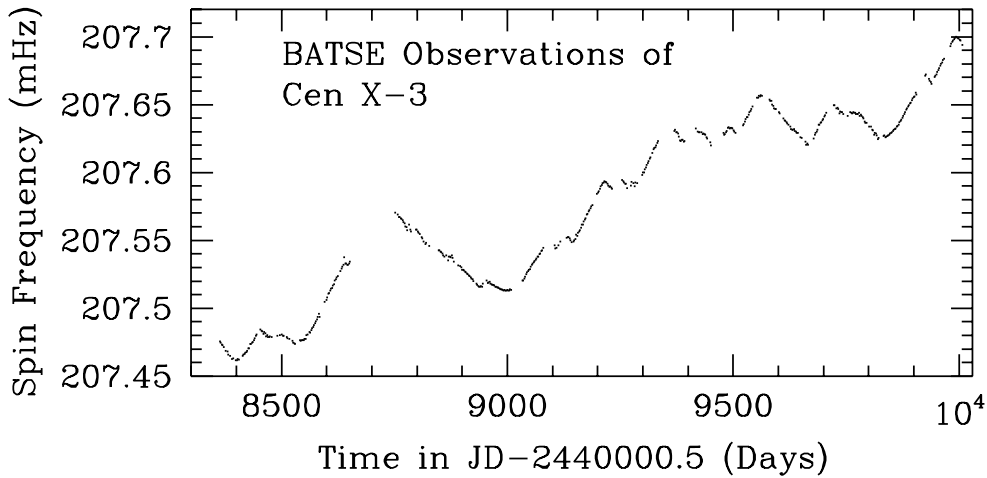,width=5.25in,height=2.55in}}
\vspace{3pt}
\caption{The intrinsic spin frequency history of the accreting X-ray
pulsar Cen X-3 from BATSE observations \protect\cite{bil97}. The orbit has
been subtracted so that these frequencies are the NS spin frequency. The negligible
error bars are omitted. The data gap near day 8700 is from a tape
recorder error, the other gaps are real non-detections of the pulsar
due to low luminosities.
}\label{fig:cenx3}
\end{figure}

The early picture of long-term pulsar spin evolution was based on
sparse measurements of $\sim 10$ objects \cite{gl79,jossrap84}.  In
particular, the spin behavior of Cen X-3 and Her X-1
suggested that the simple spin-up torque estimate in equation
(\ref{eq:torque}) was sometimes inadequate. These pulsars were
apparently spinning up on a timescale much longer than predicted by
equation (\ref{eq:maxtorque}). Moreover, both sources also underwent
short episodes of spin-down, indicating that angular momentum was
actually being lost by the pulsar while it continued to accrete.

Figure \ref{fig:cenx3} is the frequency history of the 4.8\,s pulsar
Cen X-3 from BATSE \cite{bil97}. This is an example where BATSE
observations reveal a strikingly different picture of pulsar spin
behavior than previously known \cite{nel97}. Prior to 1991, the
long-term frequency evolution was described as secular spin-up at
$\dot \nu = 8\times10^{-13}$\,Hz\,s$^{-1}$, a factor of 10-20 slower
than predicted by equation (\ref{eq:nudot}) for the roughly known
accretion rate of $\dot M\approx (3-6)\times 10^{-9} M_\odot \ {\rm
yr^{-1}}$.  In contrast, the BATSE data show that Cen X-3 exhibits
10$-$100\,d intervals of steady spin up and spin down at a much larger
rate consistent with the simple estimate of equation
(\ref{eq:maxtorque}). The torques have a bimodal distribution, with
the average spin up torque ($7\times 10^{-12} \ {\rm Hz \ s^{-1}}$),
larger than the average spin down torque ($3\times10^{-12}\ {\rm Hz \
s^{-1}}$) \cite{nel97}.  Transitions between spin up and spin down
occur on a time scale more rapid than BATSE can resolve (less than a
day). More significantly, the long-term spin-up rate inferred from the
pre-BATSE data is not representative of the instantaneous torque; its
small value is a consequence of the frequent torque transitions of
roughly $\pm N_o$.

We now know that this switching behavior is common. In 1991, the
7.6\,s pulsar 4U~1626--67 underwent a reversal to spin down at a rate
$\dot \nu=-7\times10^{-13} \ {\rm Hz \ s^{-1}}$ after 20 years of spin
up with $\dot \nu = +8.5\times10^{-13} \ {\rm Hz \ s^{-1}}$
\cite{chak97}. The final torque is nearly equal in magnitude but
opposite in sign and the magnitude is comparable to $N_o$. The pulsar
GX 1+4 underwent a similar transition to spin-down in 1988
\cite{mak88} after more than a decade of steady spin up. Again, the
spin down rate is close in magnitude to the spin up rate
\cite{chak97b} and of order $N_o$. The 38\,s pulsar OAO~1657--415 has
torque episodes similar to those seen in Cen X-3 \cite{chak93}.

 There are at least two classes of models that might explain
instantaneous spin-down in disk-fed pulsars, and both involve the
interaction between the accretion disk and the stellar magnetosphere.
Ghosh and Lamb (1979)\cite{gl79} argued that Her X-1 and Cen X-3 must
be near equilibrium and found that additional negative torques would
then act on the NS. Stars sufficiently close to equilibrium might then
spin-down while continuing to accrete. Other theories explain
spin-down via the loss of angular momentum in an MHD outflow
\cite{anz80,aro84,lov94}. Outflowing material moves along rigid
magnetic field lines like beads on a wire, gaining angular momentum as
it is forced to corotate. This can result in a spin-down torque on the
star if the matter leaving is somehow tied to field lines that
eventually connect to the star. This connection remains a challenge. 

The BATSE observations pose difficulties for all these theories. To
produce the bimodal torque behavior, all near-equilibrium models
require delicate changes in $\dot M$. The transitions in Cen X-3
always alternate between torques of opposite sign. Why should this
occur? The situation is even more challenging for 4U~1626--67, where
$\dot M$ is most likely set by the loss of orbital angular momentum
via gravitational radiation. Having the companion switch to such a
finely-tuned mass transfer rate so that the spin-down torque would
have nearly the same magnitude as the previous spin-up torque seems
difficult. These difficulties (as well as the anti-correlation of
torque and flux in GX 1+4 \cite{chak97b}) led Nelson et al. (1997)
\cite{nel97} to hypothesize that the accretion disks were alternating
their sense of rotation, becoming retrograde at times. Though a
radical hypothesis, it just might be what is needed.

\section{Weakly Magnetic Neutron Stars} 

The launch of the {\it Rossi X-Ray Timing Explorer} (RXTE) has allowed
for the discovery of fast quasi-periodic variability from many rapidly
accreting ($ \dot M > 10^{-10} M_\odot \ {\rm yr^{-1}}$) and weakly
magnetic ($B\ll 10^{11} \ {\rm G}$) NSs in low-mass X-ray binaries
\cite{vdk97}. These observations strongly suggest that these NSs are
rapidly rotating, as predicted by those scenarios connecting the
millisecond radio pulsars to this accreting population
\cite{bhat}. Discoveries have occurred on two different fronts: (1)
kHz QPO's in the persistent emission \cite{vdk97} and, (2)
periodicities during Type I bursts. I will summarize the latter here.

This began with the detection of nearly coherent $\nu_B=363$ Hz
oscillations during type I X-ray bursts from the low accretion rate
($\dot M < 10^{-9} M_\odot {\rm yr^{-1}}$) NS 4U~1728-34
\cite{stroh96}. Pulsations with amplitudes of $2.5-10 \%$ were
detected in six of the eight bursts analyzed at that time. In
addition, two separate high frequency quasi-periodic oscillations
(QPOs) were found in the persistent emission. These changed with
accretion rate, but maintained a fixed difference frequency of
$\nu_d\approx 363 $ Hz, identical to the period seen during the
bursts.  The detection of two drifting QPO's (in the persistent
emission) separated by a fixed frequency identical to that seen in the
bursts naturally leads to beat frequency models
\cite{stroh96,miller}. The difference frequency is presumed to be the
NS spin frequency, $\nu_s=1/P_s$, whereas the upper frequency has different
origins in different models (see \cite{vdk97} for a summary). Most importantly, we
expect to see the spin during the burst rise, as it is during this
time that the nuclear burning front is engulfing the whole star,
leading to a temporarily large brightness asymmetry on the star
\cite{fry82,noz84,bil93,bil95}. The temporal behavior of the periodic
oscillation during the rise of the bursts is consistent with this
explanation \cite{stroh97b}.

  There are presently six NSs with measured periodicities during Type
I X-ray bursts. The burst frequencies, $\nu_B$ and kHz QPO difference
frequencies, $\nu_d$, are as follows: 4U~1702-429 ($\nu_B=330$ Hz,
\cite{swank}), 4U~1728-34 ($\nu_B=\nu_d=363$ Hz, \cite{stroh96}),
KS~1731-260 ($\nu_B=524$ Hz, $\nu_d=260\pm 10$ Hz,
\cite{smith,wijvdk}), Aql~X-1 ($\nu_B=549 $ Hz, \cite{zhan98}),
4U~1636-53 ($\nu_B= 581$ Hz, $\nu_d=276\pm 10$ Hz,
\cite{zhan97a,wijnands}), MXB~1743-29 ($\nu_B= 589 $ Hz
\cite{stroh97a}).  Clearly, both the difference frequencies and the
measured frequencies in the Type I bursts are in a rather narrow
range, from 260 to 589 Hz.  Indeed, if some of the frequencies seen
during the Type I bursts are doubled due to, for example weak dipolar
fields, then the inferred spin frequencies are all within an even
narrower range (262-363 Hz). There are also many NSs that accrete at
higher rates and are not regular Type I X-ray bursters. Many of these
objects, notably the ``Z'' sources, also show drifting QPO's at fixed
separation, again with a similarly narrow frequency range (250-350
Hz). Beat-frequency like models are also applied to these observations
so as to infer $\nu_s$.\footnote{The applicability of such a model is
less clear when the difference frequency is not constant, as in Sco
X-1 \cite{van97} and 4U 1608-52 \cite{mend98}. }

 The frequency observed in the cooling tail of bursts from MXB
1743-29\cite{stroh97a}, 4U 1728-34\cite{stroh97b} and Aql
X-1\cite{zhan98} increased by a few Hz as the flux decreased. There is
no torque large enough to change the neutron star spin this rapidly,
so Strohmayer et al. (1997) \cite{stroh97a} argued that the observed
periodicity is the rotation rate of the burning shell alone. They
noted that the slight ($\Delta r \ll R$) hydrostatic expansion during
the burning can explain the observations if the shell conserves
angular momentum. In that case, the change in thickness (roughly
$\Delta r \approx 20$ meters, \cite{bil95}) will lead to a frequency
change of the burning material by an amount $\Delta \nu\approx \nu_s
(\Delta r/R)\approx 2\times 10^{-3} \nu_s $, or a change of 1 Hz for a
500 Hz rotation. This is close to what is observed. It is presumed
that the neutron star spin frequency is the higher value and is 
unchanging throughout the burst. Indeed, well separated
observations of 4U 1728-34 find the same asymptotic frequency
\cite{stroh96,stroh97b}.
 
 Why is the observed frequency evolution always from low to high
during the cooling tail? It is most likely because the atmospheric
scale height decreases during the cooling phase \cite{stroh97a}. The
atmosphere expanded and spun-down at the onset of the thermonuclear
instability. However the radiative layer on the neutron star delays
the information about the burst. Hence, the small amount of
hydrostatic expansion and spin-down has already occurred by the time
the observer sees the burst. Observing the spin-down will be difficult.
 
Allowing the burning shell to have its own rotation rate leads to
matter wrapping around the star many times during the instability,
possibly leading to a different spreading speed in the longitudinal
direction at fixed latitudes. If true, one might hope to see a
different form of burst rise depending on the rotation rate. Such
shear layers can be unstable to the Kelvin-Helmholtz
instability. However, the buoyancy due to the mean molecular weight
contrast or even thermal effects can easily stabilize this shear as
the Richardson number ${\it Ri}=N_{BV}^2/(dv_{rot}/dz)^2\gg 1/4$,
where $N_{BV}> 10 \ {\rm kHz}$ is the local Brunt-V\"ais\"al\"a
buoyant frequency \cite{bilcum}. The shear might thus persist for the
few seconds required. Further theoretical studies need to be carried
out to fully understand the repercussions of this result. Problems
that immediately come to mind are the role of any magnetic field and
more importantly, the possibility of longer timescale rotational
instabilities.

\section{Conclusions and Open Questions} 

\begin{figure}
\centerline{\epsfig{file=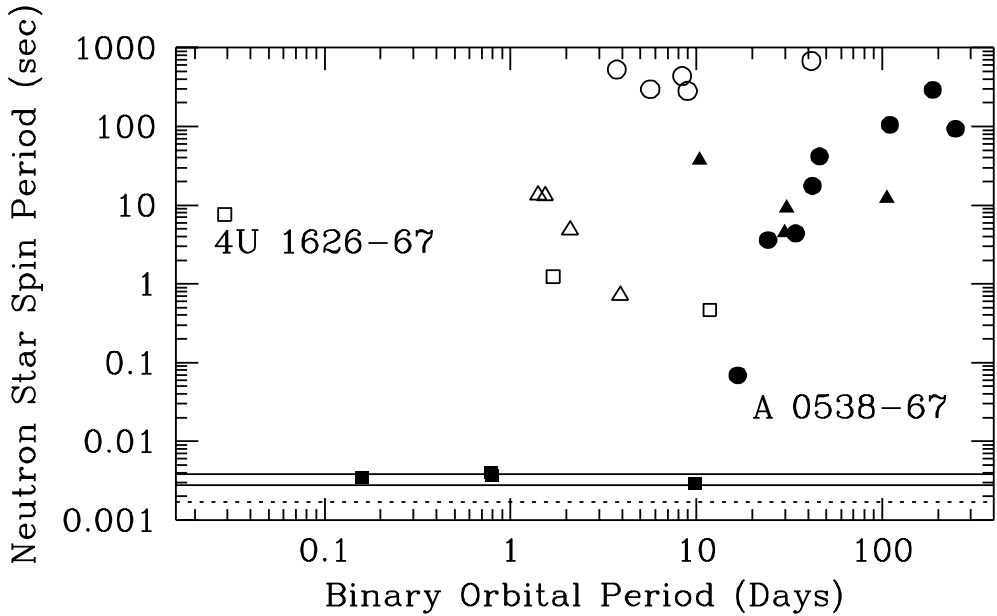,width=5.25in,height=3.25in}}
\vspace{3pt}
\caption{The NS spin period versus the binary orbital period (the
``Corbet'' diagram,\protect\cite{corbet}). The X-ray pulsar data are
from Table 1 of \protect\cite{bil97}. The symbols for the X-ray
pulsars are: Be-type binaries (filled circles), wind-fed massive
binaries (open circles), Roche-lobe overflow fed massive binaries
(open triangles), unknown companion (filled triangles), Roche-lobe fed
low-mass binaries (open squares). The filled squares are the four
LMXB's (Sco X-1, Cyg X-2, 4U 1636-53 \& Aql X-1) for which the kHz QPO
inferred spin period \protect\cite{vdk97} and orbital period
\protect\cite{van95} are known. The two solid horizontal lines show
the range of inferred spin frequencies from the kHz QPO's alone
\protect\cite{vdk97} (i.e. $\nu_d$).  This would be the range of
inferred $\nu_s$ if there is frequency doubling during Type I
bursts. If the burst frequencies are $\nu_s$, then the dashed line
represents the shortest period yet seen in an accreting NS (1.69 ms
\protect\cite{stroh97a}).
\label{corbetfig}}
\end{figure}

 So, where do things stand?  Figure \ref{corbetfig} displays the spin
periods and orbital periods for all accreting NS where both have been
measured. As is evident, the X-ray pulsars (all objects with $P_s> 10
$ ms on this plot) have a large range of spin periods, presumably due
to the large range of magnetic field strengths and accretion rates.
The outliers are identified by name. Clearly the most rapidly rotating
X-ray pulsar A 0538-67 needs to be confirmed.  What is most striking
about this figure is the narrow range of frequencies where the LMXB's
reside (see caption for details). It is remarkable that these stars
are all rotating at nearly the same rate. White and Zhang (1997)
\cite{white} argued that this similarity arises because these weakly
magnetic ($B\sim 10^8-10^9 \ {\rm G}$) NSs have reached an equilibrium
where the magnetospheric radius equals the co-rotation radius. The NSs
must then have an intrinsic relation between their magnetic fields and
accretion rates so that they all reach a rotational equilibrium of
roughly the same frequency.  The mapping of this onto the NS
parameters is a bit uncertain for magnetospheres this close to the
NS. Naively applying the standard scalings would imply that
$\mu\propto \dot M^{1/2}$ for this to be true. Clearly, nothing like
this is seen for the highly magnetic X-ray pulsars. The similarities
in spin frequencies remains to be explained and is also motivated by
Backer's \cite{backer} recent claim of a similarly narrow range of
spin periods at birth for the millisecond radio pulsars.

What about the magnetic X-ray pulsars? The BATSE observations suggest
that the disk-accreting pulsars are always subject to instantaneous
torques of magnitude $\approx N_o\equiv \dot M (GMr_{co})^{1/2}$ and
only differentiate themselves by the timescale for reversals of
sign. We see some (e.g. Cen X-3, Her X-1, OAO 1657-415) that switch
within $\sim 10-90$ days, whereas others (e.g. 4U 1626-67 and GX 1+4)
switch once in 10-20 years. The primary theoretical issues are then
identifying the physics that sets this timescale and understanding why
the magnitudes of the spin-up and spin-down torques are the
same. Whether this is because these objects are at, or near, the
equilibrium spin period is still an open question.

It is intriguing to apply this picture of the long-term evolution of
disk-fed pulsars to those that are not monitored by BATSE. First, it
makes it more plausible that one of the class of the ``6 second''
pulsars which are spinning down (1E 2259+586, 1E 1048.1-5937, 4U
0142+61,\cite{mer}) might eventually switch to spin-up. The pulsar 1E
1048.1-5937 has the shortest spin-down time amongst these
($t_{sd}=10^4 \ {\rm yr}$) and might be the most likely one to undergo
a torque reversal. There is already some evidence for a brief torque
reversal in 1E 2259+586 \cite{baykal}. In addition, the long-term
torque inferred for LMC X-4 is nearly a factor of 100 lower than
$N_o$, suggesting that this pulsar may be undergoing rapid switching
like Cen X-3. Repeated torque measurements of this object should bear
this out. 

I thank Tod Strohmayer and Michiel van der Klis for continuous
information on the spin periods during Type I bursts and the kHz
QPOs. I also thank Ed Brown, Deepto Chakrabarty, Mark Finger, Rob
Nelson, Tom Prince and Brian Vaughan for countless discussions
concerning X-ray pulsars. This work was supported by NASA grant
NAGW-4517 and the Alfred P. Sloan Foundation.

\end{document}